# The Earth-Moon system as a typical binary in the Solar System
## S. I. Ipatov (1,2)

(1) Vernadsky Institute of Geochemistry and Analytical Chemistry of Russian Academy of Sciences, Kosygina 19, 119991, Moscow, Russia; (2) Space Research Institute of Russian Academy of Sciences, Profsoyuznaya st. 84/32, Moscow, Russia (siipatov@hotmail.com).

**Abstract**
Solid embryos of the Earth and the Moon, as well as trans-Neptunian binaries, could form as a result of contraction of the rarefied condensation which was parental for a binary. The angular momentum of the condensation needed for formation of a satellite system could be mainly acquired at the collision of two rarefied condensations at which the parental condensation formed. The minimum value of the mass of the parental condensation for the Earth-Moon system could be about 0.02 of the Earth mass. Besides the main collision, which was followed by formation of the condensation that was a parent for the embryos of the Earth and the Moon, there could be another main collision of the parental condensation with another condensation. The second main collision (or a series of similar collisions) could change the tilt of the Earth. Depending on eccentricities of the planetesimals that collided with the embryos, the Moon could acquire 0.04-0.3 of its mass at the stage of accumulation of solid bodies while the mass of the growing Earth increased by a factor of ten.

**Introduction**

It is supposed by many authors that the Earth-Moon system formed as a result of a collision of the solid Earth with a Mars-sized object. Galimov and Krivtsov [1] presented arguments that such giant impact concept has several weaknesses. Below we discuss that the formation of the Earth-Moon system from a rarefied condensation can be considered similar to formation of trans-Neptunian binaries. Ipatov [2-3] and Nesvorny et al. [4] supposed that trans-Neptunian satellite systems formed by contraction of rarefied condensations. Ipatov [3] concluded that trans-Neptunian satellite systems could get the main fraction of their angular momenta due to collisions of rarefied condensations.

Last years, new arguments in favor of the formation of rarefied preplanetesimals (clumps) were presented in several papers. These clumps could include decimeter and meter sized boulders in contrast to dust condensations earlier considered. For example, Johansen et al. [5] found the efficient formation of gravitationally bound clumps, with a range of masses corresponded to radii of contracted objects from 100 to 400 km in the asteroid belt and from 150 to 730 km in the Kuiper belt. Lyra et al. [6] showed that in the vortices launched by the Rossby wave instability in the borders of the dead zone, the solids quickly achieve critical densities and undergo gravitational collapse into protoplanetary embryos in the mass range [$0.1M_E$, $0.6M_E$], where $M_E$ is the mass of the Earth.

**Initial Angular Velocities of Rarefied Condensations and Angular Velocities Needed for Formation of Satellite Systems**

According to Safronov [7], the initial angular velocity of a rarefied condensation (around its center of mass) was $0.2\Omega$ for a spherical condensation, where $\Omega$ is the angular velocity of the condensation moving around the Sun. The initial angular velocity is positive and is not enough for formation of satellites. In calculations of contraction of condensations (of mass $m$ and radius $r=0.6r_H$, where $r_H$ is the Hill radius) presented in [4], trans-Neptunian objects with satellites formed at initial angular velocities $\omega_o$ from the range [$0.5\Omega_o$, $0.75\Omega_o$], where $\Omega_o=(Gm/r^3)^{1/2}$, $G$ is the gravitational constant. As $\Omega_o/\Omega \approx 1.73(r_H/r)^{3/2}$, then $\Omega \approx 0.27\Omega_o$ and $0.2\Omega \approx 0.054\Omega_o$ at $r=0.6r_H$.

In the 3D calculations of gravitational collapse of the parental condensation for the Earth-Moon system presented in [1], binaries formed at $\omega_o/\Omega_o$ from the range [1, 1.46]. The radii of initial condensations used in calculations considered in [1] were much smaller (by about a factor of 40) than their Hill radii. Galimov and Krivtsov [1] considered evaporation of particles that constitute a rarefied condensation in order to explain the formation of the Earth-Moon system from the condensation with the same angular momentum as that of this system. May be such formation could take place in the model without the evaporation if one consider another size of the parental condensation in their calculations? It may be interesting to calculate the contraction of condensations for a wider range of ratios of radii of condensations to their Hill radii than in the calculations presented in [1,4].

**Formation of Trans-Neptunian Binaries**

Nesvorny et al. [4] calculated contraction of rarefied condensations in the trans-Neptunian region and found the cases when the contraction ends in formation of binaries (or triples). They supposed that condensations got their angular momenta when they formed from the protoplanet cloud. Ipatov [3] compared the angular velocities used by Nesvorny et al. [4] as initial data for contracting preplanetesimals with the angular velocity of the condensations formed by the merger between two collided condensations which moved before the collision in circular heliocentric orbits. The angular velocity ω of the condensation formed at the collision of two identical condensations moving in circular heliocentric orbits can be as high as $1.575\Omega$ [3]. At $r=r_H$, ω can be as high as $0.9\Omega_o$, or even a little greater than $\Omega_o$, if we take into account the initial $0.2\Omega$. Ipatov concluded [3] that the initial angular velocity of the parental condensation at which a binary could form at calculations made by Nesvorny et al. [4] could be obtained at the collision of two condensations which moved before the collision in circular heliocentric orbits. For the case when sizes of collided rarefied preplanetesimals (RPPs) are equal to the sizes of RPPs at the moment when they got initial rotation, the typical angular momentum acquired at the collision of two equal condensations is greater by an order of magnitude than the initial angular momentum of the condensation with mass equal to that of the parental condensation.

The role of initial rotation in the angular momentum $K_s$ of the parental RPP can be greater if sizes of RPPs changed before the collision. If we consider a collision of two identical spherical RPPs with masses $m_1$ and radii equal to $k_{col}r_H$, each of which initially formed with radius $k_{in}r_H$ and angular velocity equal to $0.2\Omega$, then the angular momentum of the spherical RPP formed after the collision is $K_s \approx (0.96 k_\Theta \cdot k_{col}^2 + 0.077 \cdot k_{in}^2) a^{1/2} m_1^{5/3} G^{1/2} M_S^{-1/6}$, where $a$ is the semi-major axis of the RPP, and $M_S$ is the mass of the Sun. According to [2], the mean value of $k_\Theta$ is 0.6 and $k_\Theta \le 1$. In the above formula at $k_{in}/k_{col}>2.7$ and $k_\Theta=0.6$ (or at $k_{in}/k_{col}>3.5$ and $k_\Theta=1$), the role of initial rotation is greater than the role of the collision. It shows that collisions played the main role in $K_s$ only when sizes of preplanetesimals did not differ much (by not more than a factor of 3) from their initial sizes. At $k_{in}/k_{col}$ about 3, masses differ by a factor of about 30. If we consider formation of a condensation at a collision of two RPPs for which the ratio of their radii is equal to $k_r$, then at $k_\Theta=1$ and $k_{in}=k_{col}$, the contribution to $K_s$ due to a collision of the RPPs is greater than the contribution to $K_s$ due to initial rotation (equal to $0.2\Omega$) by a factor of 12.5, 3 and 0.8 for $k_r$ equal to 1, 2, and 3, respectively. Below we use the term 'similar sizes of RPPs' for the case when the collision of considered RPPs gives the main contribution to the angular momentum of the final RPP, i.e., the ratio of diameters of collided RPPs does not exceed 3 for almost circular heliocentric orbits. At some collisions, the mass of the formed RPP could be smaller than the sum of masses of collided RPPs, and some fraction of the angular momentum of collided RPPs could belong to the material that was not incorporated into the formed RPP.

**Prograde and Retrograde Rotation of Trans-Neptunian Binaries**

Ipatov [7] studied inclinations $i_s$ of orbits of secondaries around 32 objects moving in the trans-Neptunian belt and discussed how such inclinations could form. The below discussion is based on the plots presented in [7], and the plots used the data from http://www.johnstonsarchive.net/astro/astmoons/. Note that $i_s$ is considered relative to the ecliptic and differs from the inclination relative to the plane which is perpendicular to the axis of rotation of a primary. For example, $i_s=96°$ for Pluto, though Charon is moving in the plane which is perpendicular to the Pluto's rotational axis. The fraction of objects with $i_s>90°$ equals $13/32\approx0.406$ at all values of eccentricity $e$ of a heliocentric orbit of a binary, and it is $13/28\approx0.464$ for $e<0.3$. The distribution of $i_s$ is in the wide range almost from 0 to 180°. It shows that a considerable fraction of the angular momentum of the RPPs that contracted to form satellite trans-Neptunian systems was not due to initial rotation of RPPs or to collisions of RPPs with small objects (e.g., boulders and dust), but it was acquired at collisions of the RPPs which masses did not differ much, because else the angular momentum would be positive. Ipatov [2] noted that the angular momentum of collided RPPs could be positive or negative depending on heliocentric orbits of the RPPs. Some excess of the number of discovered binaries with positive angular momentum compared with the number of discovered binaries with negative angular momentum was caused in particular by the contribution of initial positive angular momentum of RPPs and by the contribution of collisions of RPPs with small objects to the angular momentum of the parental RPP that produced the binary. There could be also some excess of positive angular momentum at mutual collisions of RPPs of similar sizes.

We suppose that in the case of two centers of contraction inside the preplanetesimal formed by a merger between two collided rarefied condensations, at the values of the angular momentum of the parental preplanetesimal a little smaller than it is needed for formation of binaries, a solid object consisting of two touching components could form. Some asteroids and comets (e.g., asteroid Itokava and Comet 67P/Churyumov–Gerasimenko) consist of two parts and have a form of dumbbells. Probably, the mass range of initial rarefied condensations in the Solar System was very wide, and their masses could vary from a mass of 1-km solid object to the mass of Mars.

**The Angular Momentum of the Condensation Parental for the Earth-Moon System that Formed at a Collision of Two Condensations**

The angular velocity of the parental condensation formed at the collision of two identical condensations is a little smaller than $\Omega_o$ needed for formation of binaries in calculations made by Galimov and Krivtsov [1], but the contraction of the condensation formed at the collision to the size of the condensation considered in [1] can considerably increase the angular velocity. The angular velocity of the condensation of radius $r_c$ formed as a result of compression of the condensation, with radius $r_1$ and the angular velocity $\omega_1$, equals $\omega_{rc}=\omega_1(r_1/r_c)^2$. The angular momentum of the condensation of radius $0.12r_H$ formed at a typical collision of two identical condensations is the same as the angular momentum for the condensation with $r=0.025r_H$ considered in [1]. Therefore, any initial angular velocities considered in [1,4] can be reached after contraction of the condensation formed at a collision of condensations not greater than their Hill spheres.

In our opinion, the embryos of the Earth and the Moon could form as a result of contraction of the same parental rarefied condensation. A considerable fraction of the angular momentum of such condensation could be acquired at a collision of two rarefied condensations. Based on formulas presented in [2-3], we have concluded that the present angular momentum of the Earth-Moon system could be acquired at the collision of two identical rarefied condensations with sizes of Hill spheres, which total mass was about $0.1M_E$, and which heliocentric orbits were circular. The initial mass of the rarefied condensation that was a parent for the embryos of the Earth and the Moon could be relatively small ($0.02M_E$ or even less) if we take into account the growth of the angular momentum of the embryos at the time when they accumulated planetesimals. There could be also the second main collision of the parental condensation with another condensation, at which the radius of the Earth's embryo condensation was smaller than the semi-major axis of the orbit of the Moon's embryo. The second main collision (or a series of similar collisions) could change the tilt of the Earth to its present value.

**The Angular Momentum of the Rarefied Condensation that Formed by Accumulation of Smaller Objects**

If the parental condensation with final mass $m$ and radius $r=k_H r_H$ got all its angular momentum $K_s$ as a result of accumulation of smaller objects, then $K_s \approx 0.173 k_H^2 G^{1/2} a^{1/2} m^{5/3} M_S^{-1/6} \Delta K$ [3], where $\Delta K$ is the difference between the fraction of positive increments of angular momentum and the fraction of negative increments. At $\Delta K=0.9$ (a typical value for Hill spheres moving in circular heliocentric orbits), $m=M_E+M_M$ (the sum of present masses of the Earth and the Moon), $k_H=1$, and $a=1$ AU, we obtain that $K_s$ is greater by a factor of 24.5 than the present angular momentum $K_{sEM}$ of the Earth-Moon system, including the rotational momentum of the Earth. Taking into account that $K_s$ is proportional to $m^{5/3}$, we obtain that $K_s=K_{sEM}$ at $m=(M_E+M_M)/6.8$. The angular momentum of the Earth-Moon system is positive. Therefore, for the mass of the final condensation $m \geq 0.15M_E$, the angular momentum equal to $K_{sEM}$ could be acquired at any contribution of a collision of two large condensations to the angular momentum of the final condensation. In principle, the angular momentum of the condensation needed for formation of the Earth-Moon system could be acquired by accumulation only of small objects. Nevertheless, we suppose that the collision of two large condensations played a considerable role in the angular momentum of the collapsing parental condensation. Else the parental condensations of Venus and Mars could also get large angular momentum, which was enough for formation of large satellites. The greater was the role of small objects in formation of the condensation that was a parent for the Earth-Moon system, the greater could be the difference in masses of two collided condensations at the main collision. It may be a question whether two condensations which masses differed by not more than an order of magnitude could form at close distances from the Sun. Dust particles and bolders could considerably change distances from the Sun with time and could reach the growing condensation from not close distances if the lifetime of the condensation was not small. In order to get large times of contraction of condensations, it is needed to consider factors preventing fast collapse of condensations.

**The Growth of Solid Embryos of the Earth and the Moon**

Solid embryos of the Earth and the Moon grew to the present masses of the Earth and the Moon ($M_E$ and $0.0123 M_E$, respectively) by accumulation of smaller planetesimals. For large enough eccentricities of planetesimals, the effective radii $r_{ef}$ of proto-Earth and proto-Moon are proportional to the radius $r_e$ of a considered embryo. For such proportionality, we can obtain $r_{Mo} = m_{Mo}/M_E = [(0.0123^{-2/3} - k + k \cdot (m_{Eo}/M_E)^{-2/3})]^{-3/2}$, where $k = k_d^{-2/3}$, $k_d$ is the ratio of the density of the growing Moon of mass $m_M$ to that of the growing Earth of mass $m_E$ ($k_d = 0.6$ for the present Earth and Moon), $m_{Mo}$ and $m_{Eo}$ are initial values of $m_M$ and $m_E$. For $r_{Eo} = m_{Eo}/M_E = 0.1$, we have $r_{Mo} = 0.0094$ at $k = 1$ and $r_{Mo} = 0.0086$ at $k = 0.6^{-2/3}$. At these values of $r_{Mo}$, the ratio $f_m = (0.0123 - r_{Mo})/0.0123$ of the total mass of planetesimals that were accreted by the Moon at the stage of the solid body accumulation to the present mass of the Moon is 0.24 and 0.30, respectively. For $r_{ef}$ proportional to $r_e$ and the growth of the mass of the Earth embryo by a factor of ten, the mass of the Moon embryo increased by a factor of 1.31 and 1.43 at $k = 1$ and $k = 0.6^{-2/3}$, respectively.

At small relative velocities of planetesimals, effective radii of the embryos are proportional to $r_e^2$. For such proportionality, integrating $dm_M/m_M = k_2 \cdot (m_M/m_E)^{4/3} dm_E/m_E$, we can get $r_{Mo2} = m_{Mo}/M_E = [(0.0123^{-4/3} - k_2 + k_2 \cdot (m_{Eo}/M_E)^{-4/3})]^{-3/4}$, where $k_2 = k_d^{-1/3}$. For $r_{Eo} = m_{Eo}/M_E = 0.1$, we have $r_{Mo2} = 0.01178$ and $f_{m2} = (0.0123 - r_{Mo2})/0.0123 = 0.042$ at $k_2 = 1$, and $r_{Mo2} = 0.01170$ and $f_{m2} = 0.049$ at $k_2 = 0.6^{-1/3}$. In this case for the growth of the Earth embryo mass by 10 times, the Moon embryo mass increased by the factor of 1.044 and 1.051 at $k_2 = 1$ and $k_2 = 0.6^{-1/3}$, respectively. In the above models, depending on eccentricities of planetesimals (i.e., on dependence of $r_{ef}$ on $r_e$), the Moon could acquire 0.04-0.3 (the lower estimate is for almost circular heliocentric orbits) of its mass at the stage of accumulation of solid bodies during the time when the mass of the growing Earth increased by a factor of ten.

In our approach, the influx of the matter to embryos is from the zone around the heliocentric orbit of the Earth-Moon embryos system, but not only from the sphere around the embryos as in [1]. For comparison with [1], in the case of $k_d = 0.6$ and $M_E/m_{Eo} = 26.2$ we have $M_M/m_{Mo} \approx 2$ at $r_{ef}$ proportional to $r_e$ and $M_M/m_{Mo} \approx 1.19$ at $r_{ef}$ proportional to $r_e^2$, i.e., the estimates presented in [1] ($M_M/m_{Mo} \approx 1.31$) are close to our considered model at $r_{ef}$ proportional to $r_e^2$.

**Acknowledgements**

This study was supported by Program no. 9 of the Presidium of the Russian Academy of Sciences and by the Russian Foundation for Basic Research, project no. 14-02-00319.